\documentclass[aps,prl,twocolumn,showpacs,floatfix]{revtex4}
\usepackage{graphicx}
\usepackage{times}
\usepackage{nicefrac}
\usepackage{amsmath}
\usepackage{amsfonts}
\usepackage{amssymb}
\usepackage{amsthm}
\usepackage{epsf}
\usepackage{bm}
\usepackage{bbm}
\usepackage{color}

\usepackage{dcolumn}
\newcolumntype{.}{D{x}{}{-1}}

\newcommand{\be}{\begin{eqnarray}}
\newcommand{\ee}{\end{eqnarray}}

\newcommand{\rmd}{{\rm d}}

\newcommand{\rpr}{r^\prime}

\newcommand{\aZ}{\alpha Z}

\newcommand{\pd}{2p_{1/2}}
\newcommand{\pu}{2p_{3/2}}

\newcommand{\grasp}{{\textsc{grasp2k}}}
\begin{document}

\title{QED radiative corrections to the $^2$P$_{1/2}$ - $^2$P$_{3/2}$ fine-structure 
       in fluorinelike ions}

\author{A. V. Volotka$^{1,2,3}$}
\author{M. Bilal$^{1,2,4}$}
\author{R. Beerwerth$^{1,2,4}$}
\author{X. Ma$^{5}$}
\author{Th. St\"ohlker$^{1,2,6}$}
\author{S. Fritzsche$^{1,2,4}$}

\affiliation{
$^1$ Helmholtz-Institut Jena, D-07743 Jena, Germany\\
$^2$ GSI Helmholtzzentrum f\"ur Schwerionenforschung, D-64291 Darmstadt, Germany\\
$^3$ Department of Physics, St. Petersburg State University, 198504 St. Petersburg, Russia\\
$^4$ Theoretisch-Physikalisches Institut, Friedrich-Schiller-Universit\"at, D-07743 Jena, Germany\\
$^5$ Institute of Modern Physics, Chinese Academy of Sciences, 730000, Lanzhou, People's Republic of China\\
$^6$ Institut f\"ur Optik und Quantenelektronik, Friedrich-Schiller-Universit\"at, D-07743 Jena, Germany\\
}

\begin{abstract}
{\it Ab initio} calculations of QED radiative corrections to the $^2P_{1/2}$ - $^2P_{3/2}$ fine-structure transition energy are performed for selected F-like ions. These calculations are nonperturbative in $\aZ$ and include all first-order and many-electron second-order effects in $\alpha$. When compared to approximate QED computations, a notable discrepancy is found especially for F-like uranium for which the predicted self-energy contributions even differ in sign. Moreover, all deviations between theory and experiment for the $^2P_{1/2}$ - $^2P_{3/2}$ fine-structure energies of F-like ions, reported recently by Li {\it et al.}, Phys. Rev. A {\bf 98}, 020502(R) (2018), are resolved if their highly accurate, non-QED fine-structure values are combined with the QED corrections ab initially evaluated here.
\end{abstract}


\maketitle
%
%
Since quantum electrodynamics (QED) has first been worked out at the end of 1940s, this theory has been found ubiquitous for studying all phenomena associated with electrically charged particles and has meanwhile reached a high level of accuracy. The great successes of QED have also lead to its
cornerstone place in the description of other interactions in the standard model. In view of this decisive role, it is of a crucial importance to verify QED under different conditions and to set new limits for its validity. In these studies, highly charged ions provide a unique scenario for probing QED effects in the strongest electromagnetic fields accessible at present for experimental study. In recent years it was particularly shown that highly charged ions are a spectacular tool for high precision measurements. For example, it is worth to mention the experiments on binding and transition energies in few-electron heavy ions \cite{brandau:2003:073202,gumberidze:2004:203004,gumberidze:2005:223001,beiersdorfer:2005:233003,bruhns:2007:113001,thorn:2009:163001,mackel:2011:143002,amaro:2012:043005,chantler:2012:153001,gassner:2018:073033}, which provided diverse tests of the bound-state QED. In the future, the advent of new accelerator facilities, e.g., FAIR in Germany and HIAF in China, as well as storage rings and ion traps will force the limits of accuracy to ever higher and higher levels.

From a theoretical viewpoint, all tests of QED for many-electron atoms and ions are restricted by the uncertainty of the electron-electron correlation effects. Since these correlation effects rapidly increase with the number of electrons, rigorous QED calculations up to the second order in $\alpha$ have been so far performed only for ions with one, two, three, four, five, or eleven electrons (see, e.g., Refs.~\cite{soff:1988:5066,mohr:1992:4421,persson:1996:204,yerokhin:1999:3522,labzowsky:1999:2707,yerokhin:2003:073001,artemyev:2005:062104,artemyev:2007:173004,kozhedub:2010:042513,sapirstein:2011:012504,artemyev:2013:032518,malyshev:2014:062517,malyshev:2015:012514,yerokhin:2015:033103,sapirstein:2015:062508,malyshev:2017:022512,blundell:1993:1790} and references therein). For ions with more electrons, in contrast, {\it ab initio} QED calculations have been performed only in first order in $\alpha$ \cite{blundell:1993:1790,sapirstein:2002:042501,sapirstein:2003:042111,chen:2006:042510} or various QED model potentials were applied to include the QED contributions approximately \cite{welton:1948:1157,indelicato:1990:5139,pyykko:2003:1469,flambaum:2005:052115,lowe:2013:118,shabaev:2013:012513}. In Refs.~\cite{ginges:2016:052509,tupitsyn:2016:253001} it has been shown that the rigorous first-order in $\alpha$ and QED model potential calculations give similar predictions. Moreover, the QED corrections evaluated by the approximate methods unambiguously bring in agreement theoretical predictions with experimental results, e.g., in the case of boronlike argon \cite{draganic:2003:183001} or neutral gold \cite{pasteka:2017:023002}. However, despite these examples, it is still not clear how accurate the QED contributions obtained by such approaches are. In particular, in recent papers of Li {\it et al.} \cite{li:2018:R020502} and Si {\it et al.}
\cite{si:2018:012504} accurate theoretical calculations have been performed for the ground-state transition energies in F-like and Co-like ions and have shown that the correlation effects are strongly reduced by so-called Layzer quenching effect \cite{layzer:1959:271,layzer:1962:177}. Thus, the accurate account for the correlations becomes possible for these transitions and this allows one to test the QED corrections. However, the theoretical results for some F-like ions are substantially different from the experimental values, as it has been reported in Ref.~\cite{li:2018:R020502}. Moreover, a careful analysis of this discrepancy showed that it does not arise from missing correlations contributions. A particular large discrepancy arises for F-like
uranium for which different QED model computations all give similar results but which all deviate by 3-6 $\sigma$ from the experiment \cite{beiersdorfer:1998:1944}.

Here, we present {\it ab initio} calculations of the QED radiative corrections for the fine-structure $1s^2 2s^2 2p^5\,^2P_{3/2} - ^2P_{1/2}$ transition energy in F-like ions. The first-order and many-electron second-order in $\alpha$ diagrams are rigorously evaluated. In particular, we show that for F-like U$^{83+}$ ion the QED corrections obtained in our work strongly disagrees with the model-potential values. Moreover, when we employ our QED values instead of the QED results obtained in Ref.~\cite{li:2018:R020502} we find excellent agreement between theoretical
and experimental values.

%
For highly charged ions, {\it ab initio} QED calculations should be relativistic from the very beginning. For these ions, in fact, the parameter $\aZ$ ($Z$ is the nuclear charge number) cannot be utilized as an expansion parameter and, hence, the calculations must be performed to all orders
in $\aZ$. As usual, let us start from the Dirac equation with the Coulomb or effective local potential, an approach that is known also as the original or extended Furry picture. The many-electron wave functions are then constructed from the one-electron Dirac solution as the Slater determinants. In the extended Furry representation, the effective potential is the sum of the Coulomb and screening potentials. In the present study, we employ two core-Hartree-type of potentials
\be
V_{\rm scr}(r) = \alpha \int_0^\infty \rmd\rpr \frac{1}{r_>}\rho(\rpr)
\ee
as screening potentials. Here, $\rho$ is the density of the {\it core} electrons. First potential (CH1) is constructed from the density of $1s^2 2s^2$ shells electrons:
\be
\rho_{\rm CH1}(r) = \sum_{c=1s^2,2s^2} \left[G_c^2(r) + F_c^2(r)\right]\,,
\ee
while for the second one (CH2) we refer $1s^2 2s^2 \pd \pu^3$ electrons as core electrons:
\be
\rho_{\rm CH2}(r) = \sum_{c=1s^2,2s^2,\pd,\pu^3} \left[G_c^2(r) + F_c^2(r)\right]\,,
\ee
where $G_c$ and $F_c$ are the upper and lower radial components of the one-electron Dirac wave function, respectively. The normalization integral of density $\rho_{\rm CH1}$ gives four, while the integral of $\rho_{\rm CH2}$ gives eight, which corresponds to the number of core electrons
in both cases.

\begin{figure}
\includegraphics[scale=1.0]{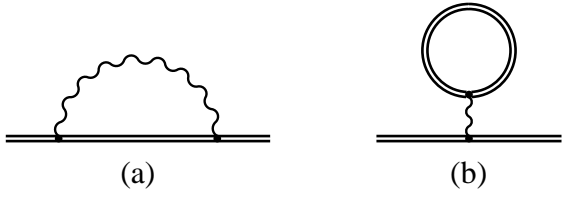}
\caption{Feynman diagrams that represent the self-energy (a) and vacuum-polarization (b) radiative corrections. The wavy line indicates the photon propagator, while the double line denotes electron propagating in the Coulomb field of a nucleus or in the effective potential.
\label{fig:1}}
\end{figure}

\begin{figure}
\includegraphics[scale=0.7]{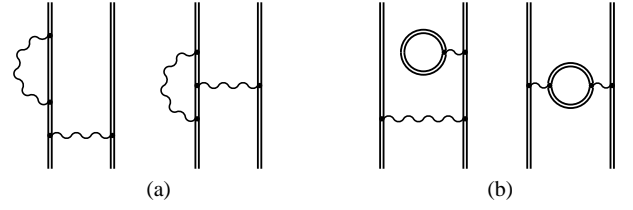}
\caption{Feynman diagrams of the screened self-energy (a) and screened vacuum-polarization (b) corrections. The notations are the same as in Fig.~\ref{fig:1}.
\label{fig:2}}
\end{figure}

Further, one has to develop the QED perturbation expansion. In the first order in $\alpha$, the radiative corrections are representing by the diagrams depicted in Fig.~\ref{fig:1}.
Their contributions to the $1s^2 2s^2 2p^5\,^2P_{3/2} - ^2P_{1/2}$ transition energy $\Delta E^{(1)}_{\rm tot}$ can be expressed as follows
\be
\label{eq:1}
\Delta E^{(1)}_{\rm tot} = \Delta E^{(1)}_{\pu} - \Delta E^{(1)}_{\pd}\,,
\ee
where $\Delta E^{(1)}_{\pu}$ and $\Delta E^{(1)}_{\pd}$ are the one-electron QED corrections to the $\pu$ and $\pd$ states, respectively. In the next (second) order in $\alpha$, the radiative corrections are given by the one-electron two-loop diagrams and many-electron (screened) one-loop diagrams, the latter are presented in Fig.~\ref{fig:2}.
In the present investigation, we do not consider the two-loop diagrams. Their contributions estimated by the one-electron values given in Ref.~\cite{yerokhin:2015:033103} are found to be smaller than our final uncertainty. Note that for the extended Furry picture one needs to account for the so-called counterterm diagrams in addition to the diagrams depicted in Fig.~\ref{fig:2}. In these diagrams, the photon which mediates the interelectronic interaction is replaced by the interaction with the screening potential. Finally, the second-order contributions to the
$^2$P$_{1/2}$ - $^2$P$_{3/2}$ fine-structure transition energy can be combined into:
\be
\label{eq:2}
\Delta E^{(2)}_{\rm tot} &=& \Delta E^{(2)}_{(1s)\pu} - \Delta E^{(2)}_{(1s)\pd} \nonumber\\
                         &+& \Delta E^{(2)}_{(2s)\pu} - \Delta E^{(2)}_{(2s)\pd} \nonumber\\
                         &+& \Delta E^{(2)}_{(\pu)} - \Delta E^{(2)}_{(\pd)} - \Delta E^{(2)}_{(\pd)\pu}\nonumber\\
                         &-& \Delta E^{(2)}_{\pu,{\rm con}} + \Delta E^{(2)}_{\pd,{\rm con}}\,.
\ee
Here, $\Delta E^{(2)}_{(1s)\pu}$, $\Delta E^{(2)}_{(1s)\pd}$ and $\Delta E^{(2)}_{(2s)\pu}$, $\Delta E^{(2)}_{(2s)\pd}$ represent the screening effects to the radiative corrections to the $\pu$ and $\pd$ states due the interaction with $1s^2$ and $2s^2$ shells, respectively; $\Delta E^{(2)}_{(\pu)}$ and $\Delta E^{(2)}_{(\pd)}$ correspond to the screening corrections due the filled $\pu^4$ and $\pd^2$ shells, respectively; while $\Delta E^{(2)}_{(\pd)\pu}$ is the $\pd^2$ screening effect to the radiative correction of the $\pu$ state. Last line of Eq.~(\ref{eq:2}) displays the counterterms, which arise in the extended Furry representation. The detailed formulas for all the contributions listed in Eqs.~(\ref{eq:1}) and (\ref{eq:2}) are presented, e.g., in Refs.~\cite{yerokhin:1999:3522,kozhedub:2010:042513,malyshev:2017:022512},
and need not to be given here, nor do we need to discuss the renormalization of these diagrams, cf. Refs.~\cite{soff:1988:5066,mohr:1992:4421,persson:1996:204,yerokhin:1999:3522,labzowsky:1999:2707,yerokhin:2003:073001,artemyev:2005:062104,artemyev:2007:173004,kozhedub:2010:042513,sapirstein:2011:012504,artemyev:2013:032518,malyshev:2014:062517,malyshev:2015:012514,yerokhin:2015:033103,sapirstein:2015:062508,malyshev:2017:022512,blundell:1993:1790} and references therein.

All diagrams are evaluated numerically by making use of the dual-kinetic-balance finite basis set method \cite{shabaev:2004:130405} with basis functions constructed from $B$ splines \cite{sapirstein:1996:5213}. These calculations involve an infinite summation over the angular-momentum quantum number $\kappa$ of intermediate states. This summation has been extended up to $|\kappa_{\rm max}| = 12$ and the remaining part of the sum is estimated by a least-square inverse-polynomial fitting.

%
The QED and screened QED radiative corrections given by Eqs.~(\ref{eq:1}) and (\ref{eq:2}) have been calculated for $^2P_{3/2} - ^2P_{1/2}$ fine-structure splitting of F-like ions. For Mo$^{33+}$ and U$^{83+}$ ions, Table~\ref{tab:QED} presents the results for the self-energy and vacuum
polarization corrections separately for three different cases: Coulomb potential and two core-Hartree potentials described above.
\begin{table}
\caption{QED ($\Delta E^{(1)}_{\rm tot}$) and screened QED ($\Delta E^{(2)}_{\rm tot}$) radiative corrections to the $^2P_{3/2} - ^2P_{1/2}$ transition energy in F-like Mo$^{33+}$ ($Z = 42$) and U$^{83+}$ ($Z = 92$) ions. The self-energy and vacuum polarization contributions are presented
separately. Results are shown for three different starting potentials: Coulomb, CH1, and CH2. Our final values (Final) are compared with the results of the approximate treatments (\grasp, Welton, MQED) evaluated in Ref.~\cite{li:2018:R020502}. All values are in eV.}
\label{tab:QED}
\tabcolsep1pt
\begin{tabular}{lrrrcrrr}                                                                     \hline\hline
                          & \multicolumn{3}{c}{$Z = 42$}   && \multicolumn{3}{c}{$Z = 92$} \\ \cline{2-4}\cline{6-8}
Contr.                    & Coulomb  &   CH1    &   CH2    && Coulomb  &   CH1    &   CH2  \\ \hline
                          &                \multicolumn{7}{c}{Self-energy}                 \\
$\Delta E^{(1)}_{\rm tot}$&   0.2714 &   0.2272 &   0.2059 &&$-$0.658  &$-$0.396  &$-$0.423\\
$\Delta E^{(2)}_{\rm tot}$&$-$0.0588 &$-$0.0112 &   0.0120 &&   1.271  &   0.906  &   0.920\\
Final                     &   0.2126 &   0.2160 &   0.2179 &&   0.613  &   0.510  &   0.497\\
\grasp                    &   0.1987 &          &          &&$-$0.821  &          &        \\
Welton                    &   0.1967 &          &          &&$-$0.668  &          &        \\
MQED                      &   0.2248 &          &          &&$-$0.361  &          &        \\
                          &                \multicolumn{7}{c}{Vacuum polarization}         \\
$\Delta E^{(1)}_{\rm tot}$&   0.0079 &   0.0062 &   0.0055 &&   2.596  &   2.290  &   2.190\\
$\Delta E^{(2)}_{\rm tot}$&$-$0.0043 &$-$0.0021 &$-$0.0013 &&$-$0.680  &$-$0.330  &$-$0.221\\
Final                     &   0.0036 &   0.0041 &   0.0042 &&   1.916  &   1.960  &   1.969\\
\grasp                    &   0.0046 &          &    &\;\;\;&   2.148  &          &        \\ \hline\hline
\end{tabular}
\end{table}
As seen from this table, the employment of the screening potential decreases (screens) the first-order contributions. This reduction happened due to partially account for the next order (e.g., second order) diagrams, accelerates the convergence of the higher-order screening corrections. Moreover, it also illustrates the main idea of approximate methods to screen the hydrogenic QED results in a way that the next order many-electron screened corrections, which are absent in the approximate treatments, become negligible. In Table~\ref{tab:QED} we also compare our final results for the self-energy and vacuum polarization corrections with values obtained within the approximate treatments in Ref.~\cite{li:2018:R020502}. In particular, there are several approaches: ``\grasp'' - the standard approach in \grasp\,\cite{joensson:2013:2197}, ``Welton'' - the Welton interpretation \cite{welton:1948:1157} of the self-energy implemented by Lowe {\it et al.} \cite{lowe:2013:118}, and ``MQED'' - a model QED operator approach \cite{shabaev:2013:012513}. The comparison reveals a reasonable agreement for the case of Mo$^{33+}$. For F-like U$^{83+}$ ion, however, the vacuum polarization corrections are close, while the self-energy values, evaluated with any approximate methods, differ even in sign. The reason for this deviation is the partial cancellation of the two (first-order) terms  $\Delta E^{(1)}_{\pu}$ and $\Delta E^{(1)}_{\pd}$. This cancellation leads to a situation when the many-electron screening effects $\Delta E^{(2)}_{\rm tot}$ absent in the approximate treatments start to be dominant. In order to estimate an accuracy, the spread of the final results through starting potentials is employed. In addition, we use the scaling of the higher-order screened diagrams $(\alpha/8\pi) (\aZ)^4/Z^2$ multiplied by a factor 5. The final uncertainty is taken as a maximum of those two. We note here, that the calculations have been also performed with the Kohn-Sham local potential \cite{kohn:1965:A1133}. However, due to the marginal difference with the CH2 results, we do not tabulate those values.

\begin{figure}[!htb]
\centerline{
\resizebox{0.5\textwidth}{!}{%
  \includegraphics{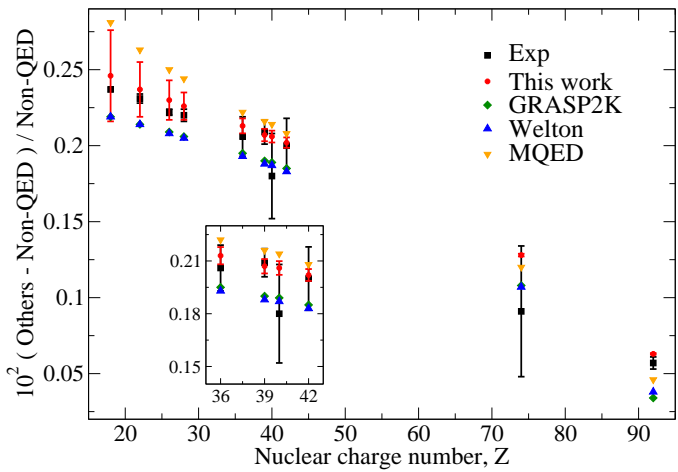}
}}
\caption{(color online). Theoretical and experimental relative values of the QED radiative corrections to the fine-structure $^2P_{3/2} - ^2P_{1/2}$ transition energy in F-like ions. The rigorous calculations of this work (red $\circ$) are compared with other theories: \grasp\,(green $\diamond$), Welton (blue $\triangle$), and MQED (orange $\triangledown$) taken from Ref.~\cite{li:2018:R020502} as well as with experimentally deduced QED values (black $\square$).
\label{fig:3}}
\end{figure}

We can also compare our values with experiments and other approximate results. In Table~\ref{tab:exp} the measured fine-structure transition energies are presented in the second column.
\begin{table*}
\caption{Comparison of the QED radiative corrections evaluated here (This work) and in Ref.~\cite{li:2018:R020502} (\grasp, Welton, MQED) with the QED corrections deduced from the experiments (Exp). The latter values are obtained as the difference between the experimental results (Experiment, 2nd column) and theoretical values without QED effects (Non-QED, 4th column) calculated in Ref.~\cite{li:2018:R020502}. All values are in eV.}
\label{tab:exp}
\tabcolsep5pt
\begin{tabular}{cr@{}lcr@{}llllll}                                                                                \hline\hline
   &    &                           &          &    &     &              \multicolumn{5}{c}{QED}               \\ \cline{7-11}
$Z$& \multicolumn{2}{c}{Experiment} & Ref.     &\multicolumn{2}{c}{Non-QED \cite{li:2018:R020502}}
                                                          & Exp       & This work  & \grasp  & Welton & MQED   \\ \hline
 18&   2&.240084(1)&\cite{draganic:2003:183001}&   2&.2348& 0.0053    & 0.0055(7)  & 0.0049  & 0.0049 & 0.0063 \\
 22&   5&.85434(19)&\cite{peacock:1984:1984}   &   5&.8408& 0.0135(2) & 0.0139(10) & 0.0125  & 0.0125 & 0.0154 \\
 26&  12&.71818(25)&\cite{peacock:1984:1984}   &  12&.6899& 0.0282(3) & 0.0292(16) & 0.0266  & 0.0265 & 0.0318 \\
 28&  17&.8486(7)&\cite{peacock:1984:1984}     &  17&.8093& 0.0393(7) & 0.0404(16) & 0.0368  & 0.0366 & 0.0435 \\
 36&  55&.352(7)&\cite{denne:1989:1488}        &  55&.238 & 0.114(7)  & 0.118(3)   & 0.108   & 0.107  & 0.123  \\
 39&  79&.071(6)&\cite{silwal:2017:30}         &  78&.906 & 0.165(6)  & 0.164(3)   & 0.150   & 0.149  & 0.171  \\
 40&  88&.465(25)&\cite{myrnaes:1994:429}      &  88&.306 & 0.159(25) & 0.182(3)   & 0.167   & 0.165  & 0.189  \\
 42& 109&.888(20)&\cite{myrnaes:1994:429}      & 109&.668 & 0.220(20) & 0.222(4)   & 0.203   & 0.201  & 0.229  \\
 74&1388&.87(60)&\cite{clementson:2011:571}    &1387&.60  & 1.27(60)  & 1.78(1)    & 1.50    & 1.48   & 1.67   \\
 92&3913&.54(16)&\cite{beiersdorfer:1998:1944} &3911&.29  & 2.25(16)  & 2.47(2)    & 1.33    & 1.48   & 1.79   \\ \hline\hline
\end{tabular}
\end{table*}
The fourth column of this table gives the results of the \grasp\,calculations \cite{li:2018:R020502}, which includes all the corrections except of QED. We can also consider a difference between the experimental results and the non-QED theoretical values. If we assume that in the theoretical calculations \cite{li:2018:R020502} all the non-QED corrections are accurately taken into account, we can interpret the difference experiment - theory as the {\it experimentally} given QED contribution and keep only the experimental uncertainty in the difference. 
These experimental QED contributions are given in the fifth column of the table. Here, we should note, that these values contain not only the radiative but also the many-electron QED corrections as well as the relativistic nuclear recoil contributions, which go beyond those implemented in the \grasp. The relativistic nuclear recoil corrections suppressed by $(\aZ)^4 m/M$ (with $M$ the mass of the nucleus) are quite small, while the many-electron QED entering at order $(\aZ)^3/Z^2$ might be important. In the sixth column the QED corrections evaluated starting with the CH2 potential (see above) are presented as our final results. Its uncertainty is estimated as a difference between the values obtained in the CH1 and CH2 potentials. Comparing the experimentally deduced QED contributions with those obtained in this work one finds an excellent agreement. From this comparison, we can conclude: ({\it i})   the correlation effects are indeed strongly diminished for the case under consideration as it has been demonstrated in Ref.~\cite{li:2018:R020502}; ({\it ii})  the Breit and, in particular, the frequency-dependent Breit interaction contributions are very well taken under control in the \grasp\,calculations; ({\it iii}) the many-electron QED effects appear to be small similar as previously revealed for He- and Li-like ions \cite{mohr:2000:052501,yerokhin:2001:032109}; ({\it iv}) the disagreement between theory and experiment discovered for some ions is now resolved. This is summarized in Fig.~\ref{fig:3} which shows the difference between various fine-structure energies and accurate non-QED calculations given also in 5th - 9th columns of Table~\ref{tab:exp}
relatively to the non-QED results. Experimental results for this ratio are compared with our QED calculations as well as with the approximate results of \grasp, Welton, and MQED computations.

Let us also compare the QED radiative corrections obtained here with the results of approximate approaches. As approximate results presented in the last three columns, \grasp, Welton, and MQED, we use the data reported in Ref.~\cite{li:2018:R020502}. The comparison with the rigorous calculations reveals that the \grasp\,and Welton approaches differ by about 10\% for all ions except U$^{83+}$. The difference with the MQED results starts from 15\% for Ar$^{9+}$ and rapidly improves to 3\% for $Z = 42$. In the case of uranium ion the approximate treatments underestimate the QED correction on a level of 46\% (\grasp), 40\% (Welton), and 28\% (MQED). The reason for this is the cancellation of the first-order corrections for the $\pu$ and $\pd$ states discussed above. Moreover, we expect similar findings for the ground-state transition in B-, C-, N-, and O-like uranium ions.

%
To summarize, rigorous QED calculations have been performed for the $1s^2 2s^2 2p^5\,^2P_{3/2} - ^2P_{1/2}$ fine-structure transition energy of selected F-like ions. The radiative corrections have been evaluated in particular up to the second order of the perturbation theory in the framework of the original and extended Furry pictures. If our QED computations are combined with highly accurate but non-QED fine-structure data by Li {\it et al.} \cite{li:2018:R020502}, the discrepancy with the experimental data is resolved and significant improvement of the agreement between theoretical and experimental data has been achieved.

%
We are grateful R. Hutton for useful discussions. This work was supported by DFG (Grant No. VO1707/1-3). A.V.V. acknowledges the support of the CAS President International Fellowship Initiative (PIFI). X.M. was supported by NSFC (Grant No. 11320101003).



\begin{thebibliography}{56}
\expandafter\ifx\csname natexlab\endcsname\relax\def\natexlab#1{#1}\fi
\expandafter\ifx\csname bibnamefont\endcsname\relax
  \def\bibnamefont#1{#1}\fi
\expandafter\ifx\csname bibfnamefont\endcsname\relax
  \def\bibfnamefont#1{#1}\fi
\expandafter\ifx\csname citenamefont\endcsname\relax
  \def\citenamefont#1{#1}\fi
\expandafter\ifx\csname url\endcsname\relax
  \def\url#1{\texttt{#1}}\fi
\expandafter\ifx\csname urlprefix\endcsname\relax\def\urlprefix{URL }\fi
\providecommand{\bibinfo}[2]{#2}
\providecommand{\eprint}[2][]{\url{#2}}

\bibitem[{\citenamefont{Brandau et~al.}(2003)\citenamefont{Brandau, Kozhuharov,
  M{\"u}ller, Shi, Shippers, Bartsch, B{\"o}hm, B{\"o}hme, Hoffknecht, Knopp
  et~al.}}]{brandau:2003:073202}
\bibinfo{author}{\bibfnamefont{C.}~\bibnamefont{Brandau}},
  \bibinfo{author}{\bibfnamefont{C.}~\bibnamefont{Kozhuharov}},
  \bibinfo{author}{\bibfnamefont{A.}~\bibnamefont{M{\"u}ller}},
  \bibinfo{author}{\bibfnamefont{W.}~\bibnamefont{Shi}},
  \bibinfo{author}{\bibfnamefont{S.}~\bibnamefont{Shippers}},
  \bibinfo{author}{\bibfnamefont{T.}~\bibnamefont{Bartsch}},
  \bibinfo{author}{\bibfnamefont{S.}~\bibnamefont{B{\"o}hm}},
  \bibinfo{author}{\bibfnamefont{C.}~\bibnamefont{B{\"o}hme}},
  \bibinfo{author}{\bibfnamefont{A.}~\bibnamefont{Hoffknecht}},
  \bibinfo{author}{\bibfnamefont{H.}~\bibnamefont{Knopp}},
  \bibnamefont{et~al.}, \bibinfo{journal}{Phys. Rev. Lett.}
  \textbf{\bibinfo{volume}{91}}, \bibinfo{pages}{073202}
  (\bibinfo{year}{2003}).

\bibitem[{\citenamefont{Gumberidze et~al.}(2004)\citenamefont{Gumberidze,
  St{\"o}hlker, Bana{\'s}, Beckert, Beller, Beyer, Bosch, Cai, Hagmann,
  Kozhuharov et~al.}}]{gumberidze:2004:203004}
\bibinfo{author}{\bibfnamefont{A.}~\bibnamefont{Gumberidze}},
  \bibinfo{author}{\bibfnamefont{T.}~\bibnamefont{St{\"o}hlker}},
  \bibinfo{author}{\bibfnamefont{D.}~\bibnamefont{Bana{\'s}}},
  \bibinfo{author}{\bibfnamefont{K.}~\bibnamefont{Beckert}},
  \bibinfo{author}{\bibfnamefont{P.}~\bibnamefont{Beller}},
  \bibinfo{author}{\bibfnamefont{H.~F.} \bibnamefont{Beyer}},
  \bibinfo{author}{\bibfnamefont{F.}~\bibnamefont{Bosch}},
  \bibinfo{author}{\bibfnamefont{X.}~\bibnamefont{Cai}},
  \bibinfo{author}{\bibfnamefont{S.}~\bibnamefont{Hagmann}},
  \bibinfo{author}{\bibfnamefont{C.}~\bibnamefont{Kozhuharov}},
  \bibnamefont{et~al.}, \bibinfo{journal}{Phys. Rev. Lett.}
  \textbf{\bibinfo{volume}{92}}, \bibinfo{pages}{203004}
  (\bibinfo{year}{2004}).

\bibitem[{\citenamefont{Gumberidze et~al.}(2005)\citenamefont{Gumberidze,
  St{\"o}hlker, Bana{\'s}, Beckert, Beller, Beyer, Bosch, Hagmann, Kozhuharov,
  Liesen et~al.}}]{gumberidze:2005:223001}
\bibinfo{author}{\bibfnamefont{A.}~\bibnamefont{Gumberidze}},
  \bibinfo{author}{\bibfnamefont{T.}~\bibnamefont{St{\"o}hlker}},
  \bibinfo{author}{\bibfnamefont{D.}~\bibnamefont{Bana{\'s}}},
  \bibinfo{author}{\bibfnamefont{K.}~\bibnamefont{Beckert}},
  \bibinfo{author}{\bibfnamefont{P.}~\bibnamefont{Beller}},
  \bibinfo{author}{\bibfnamefont{H.~F.} \bibnamefont{Beyer}},
  \bibinfo{author}{\bibfnamefont{F.}~\bibnamefont{Bosch}},
  \bibinfo{author}{\bibfnamefont{S.}~\bibnamefont{Hagmann}},
  \bibinfo{author}{\bibfnamefont{C.}~\bibnamefont{Kozhuharov}},
  \bibinfo{author}{\bibfnamefont{D.}~\bibnamefont{Liesen}},
  \bibnamefont{et~al.}, \bibinfo{journal}{Phys. Rev. Lett.}
  \textbf{\bibinfo{volume}{94}}, \bibinfo{pages}{223001}
  (\bibinfo{year}{2005}).

\bibitem[{\citenamefont{Beiersdorfer et~al.}(2005)\citenamefont{Beiersdorfer,
  Chen, Thorn, and Tr{\"a}bert}}]{beiersdorfer:2005:233003}
\bibinfo{author}{\bibfnamefont{P.}~\bibnamefont{Beiersdorfer}},
  \bibinfo{author}{\bibfnamefont{H.}~\bibnamefont{Chen}},
  \bibinfo{author}{\bibfnamefont{D.~B.} \bibnamefont{Thorn}}, \bibnamefont{and}
  \bibinfo{author}{\bibfnamefont{E.}~\bibnamefont{Tr{\"a}bert}},
  \bibinfo{journal}{Phys. Rev. Lett.} \textbf{\bibinfo{volume}{95}},
  \bibinfo{pages}{233003} (\bibinfo{year}{2005}).

\bibitem[{\citenamefont{Bruhns et~al.}(2007)\citenamefont{Bruhns, Braun,
  Kubi\v{c}ek, {Crespo L\'opez-Urrutia}, and Ullrich}}]{bruhns:2007:113001}
\bibinfo{author}{\bibfnamefont{H.}~\bibnamefont{Bruhns}},
  \bibinfo{author}{\bibfnamefont{J.}~\bibnamefont{Braun}},
  \bibinfo{author}{\bibfnamefont{K.}~\bibnamefont{Kubi\v{c}ek}},
  \bibinfo{author}{\bibfnamefont{J.~R.} \bibnamefont{{Crespo
  L\'opez-Urrutia}}}, \bibnamefont{and}
  \bibinfo{author}{\bibfnamefont{J.}~\bibnamefont{Ullrich}},
  \bibinfo{journal}{Phys. Rev. Lett.} \textbf{\bibinfo{volume}{99}},
  \bibinfo{pages}{113001} (\bibinfo{year}{2007}).

\bibitem[{\citenamefont{Thorn et~al.}(2009)\citenamefont{Thorn, Gu, Brown,
  Beiersdorfer, Porter, Kilbourne, and Kelley}}]{thorn:2009:163001}
\bibinfo{author}{\bibfnamefont{D.~B.} \bibnamefont{Thorn}},
  \bibinfo{author}{\bibfnamefont{M.~F.} \bibnamefont{Gu}},
  \bibinfo{author}{\bibfnamefont{G.~V.} \bibnamefont{Brown}},
  \bibinfo{author}{\bibfnamefont{P.}~\bibnamefont{Beiersdorfer}},
  \bibinfo{author}{\bibfnamefont{F.~S.} \bibnamefont{Porter}},
  \bibinfo{author}{\bibfnamefont{C.~A.} \bibnamefont{Kilbourne}},
  \bibnamefont{and} \bibinfo{author}{\bibfnamefont{R.~L.}
  \bibnamefont{Kelley}}, \bibinfo{journal}{Phys. Rev. Lett.}
  \textbf{\bibinfo{volume}{103}}, \bibinfo{pages}{163001}
  (\bibinfo{year}{2009}).

\bibitem[{\citenamefont{M\"ackel et~al.}(2011)\citenamefont{M\"ackel,
  Klawitter, Brenner, {Crespo L\'opez-Urrutia}, and
  Ullrich}}]{mackel:2011:143002}
\bibinfo{author}{\bibfnamefont{V.}~\bibnamefont{M\"ackel}},
  \bibinfo{author}{\bibfnamefont{R.}~\bibnamefont{Klawitter}},
  \bibinfo{author}{\bibfnamefont{G.}~\bibnamefont{Brenner}},
  \bibinfo{author}{\bibfnamefont{J.~R.} \bibnamefont{{Crespo
  L\'opez-Urrutia}}}, \bibnamefont{and}
  \bibinfo{author}{\bibfnamefont{J.}~\bibnamefont{Ullrich}},
  \bibinfo{journal}{Phys. Rev. Lett.} \textbf{\bibinfo{volume}{107}},
  \bibinfo{pages}{143002} (\bibinfo{year}{2011}).

\bibitem[{\citenamefont{Amaro et~al.}(2012)\citenamefont{Amaro, Schlesser,
  Guerra, {Le Bigot}, Isac, Travers, Santos, Szabo, Gumberidze, and
  Indelicato}}]{amaro:2012:043005}
\bibinfo{author}{\bibfnamefont{P.}~\bibnamefont{Amaro}},
  \bibinfo{author}{\bibfnamefont{S.}~\bibnamefont{Schlesser}},
  \bibinfo{author}{\bibfnamefont{M.}~\bibnamefont{Guerra}},
  \bibinfo{author}{\bibfnamefont{E.-O.} \bibnamefont{{Le Bigot}}},
  \bibinfo{author}{\bibfnamefont{J.-M.} \bibnamefont{Isac}},
  \bibinfo{author}{\bibfnamefont{P.}~\bibnamefont{Travers}},
  \bibinfo{author}{\bibfnamefont{J.~P.} \bibnamefont{Santos}},
  \bibinfo{author}{\bibfnamefont{C.~I.} \bibnamefont{Szabo}},
  \bibinfo{author}{\bibfnamefont{A.}~\bibnamefont{Gumberidze}},
  \bibnamefont{and}
  \bibinfo{author}{\bibfnamefont{P.}~\bibnamefont{Indelicato}},
  \bibinfo{journal}{Phys. Rev. Lett.} \textbf{\bibinfo{volume}{109}},
  \bibinfo{pages}{043005} (\bibinfo{year}{2012}).

\bibitem[{\citenamefont{Chantler et~al.}(2012)\citenamefont{Chantler, Kinnane,
  Gillaspy, Hudson, Payne, Smale, Henins, Pomeroy, Tan, Kimpton
  et~al.}}]{chantler:2012:153001}
\bibinfo{author}{\bibfnamefont{C.~T.} \bibnamefont{Chantler}},
  \bibinfo{author}{\bibfnamefont{M.~N.} \bibnamefont{Kinnane}},
  \bibinfo{author}{\bibfnamefont{J.~D.} \bibnamefont{Gillaspy}},
  \bibinfo{author}{\bibfnamefont{L.~T.} \bibnamefont{Hudson}},
  \bibinfo{author}{\bibfnamefont{A.~T.} \bibnamefont{Payne}},
  \bibinfo{author}{\bibfnamefont{L.~F.} \bibnamefont{Smale}},
  \bibinfo{author}{\bibfnamefont{A.}~\bibnamefont{Henins}},
  \bibinfo{author}{\bibfnamefont{J.~M.} \bibnamefont{Pomeroy}},
  \bibinfo{author}{\bibfnamefont{J.~N.} \bibnamefont{Tan}},
  \bibinfo{author}{\bibfnamefont{J.~A.} \bibnamefont{Kimpton}},
  \bibnamefont{et~al.}, \bibinfo{journal}{Phys. Rev. Lett.}
  \textbf{\bibinfo{volume}{109}}, \bibinfo{pages}{153001}
  (\bibinfo{year}{2012}).

\bibitem[{\citenamefont{Gassner et~al.}(2018)\citenamefont{Gassner,
  Trassinelli, He\ss, Spillmann, Bana{\'s}, Blumenhagen, Bosch, Brandau, Chen,
  Dimopoulou et~al.}}]{gassner:2018:073033}
\bibinfo{author}{\bibfnamefont{T.}~\bibnamefont{Gassner}},
  \bibinfo{author}{\bibfnamefont{M.}~\bibnamefont{Trassinelli}},
  \bibinfo{author}{\bibfnamefont{R.}~\bibnamefont{He\ss}},
  \bibinfo{author}{\bibfnamefont{U.}~\bibnamefont{Spillmann}},
  \bibinfo{author}{\bibfnamefont{D.}~\bibnamefont{Bana{\'s}}},
  \bibinfo{author}{\bibfnamefont{K.-H.} \bibnamefont{Blumenhagen}},
  \bibinfo{author}{\bibfnamefont{F.}~\bibnamefont{Bosch}},
  \bibinfo{author}{\bibfnamefont{C.}~\bibnamefont{Brandau}},
  \bibinfo{author}{\bibfnamefont{W.}~\bibnamefont{Chen}},
  \bibinfo{author}{\bibfnamefont{C.}~\bibnamefont{Dimopoulou}},
  \bibnamefont{et~al.}, \bibinfo{journal}{New J. Phys.}
  \textbf{\bibinfo{volume}{20}}, \bibinfo{pages}{073033}
  (\bibinfo{year}{2018}).

\bibitem[{\citenamefont{Soff and Mohr}(1988)}]{soff:1988:5066}
\bibinfo{author}{\bibfnamefont{G.}~\bibnamefont{Soff}} \bibnamefont{and}
  \bibinfo{author}{\bibfnamefont{P.~J.} \bibnamefont{Mohr}},
  \bibinfo{journal}{Phys. Rev. A} \textbf{\bibinfo{volume}{38}},
  \bibinfo{pages}{5066} (\bibinfo{year}{1988}).

\bibitem[{\citenamefont{Mohr}(1992)}]{mohr:1992:4421}
\bibinfo{author}{\bibfnamefont{P.~J.} \bibnamefont{Mohr}},
  \bibinfo{journal}{Phys. Rev. A} \textbf{\bibinfo{volume}{46}},
  \bibinfo{pages}{4421} (\bibinfo{year}{1992}).

\bibitem[{\citenamefont{Persson et~al.}(1996)\citenamefont{Persson, Salomonson,
  Sunnergren, and Lindgren}}]{persson:1996:204}
\bibinfo{author}{\bibfnamefont{H.}~\bibnamefont{Persson}},
  \bibinfo{author}{\bibfnamefont{S.}~\bibnamefont{Salomonson}},
  \bibinfo{author}{\bibfnamefont{P.}~\bibnamefont{Sunnergren}},
  \bibnamefont{and} \bibinfo{author}{\bibfnamefont{I.}~\bibnamefont{Lindgren}},
  \bibinfo{journal}{Phys. Rev. Lett.} \textbf{\bibinfo{volume}{76}},
  \bibinfo{pages}{204} (\bibinfo{year}{1996}).

\bibitem[{\citenamefont{Yerokhin et~al.}(1999)\citenamefont{Yerokhin, Artemyev,
  Beier, Plunien, Shabaev, and Soff}}]{yerokhin:1999:3522}
\bibinfo{author}{\bibfnamefont{V.~A.} \bibnamefont{Yerokhin}},
  \bibinfo{author}{\bibfnamefont{A.~N.} \bibnamefont{Artemyev}},
  \bibinfo{author}{\bibfnamefont{T.}~\bibnamefont{Beier}},
  \bibinfo{author}{\bibfnamefont{G.}~\bibnamefont{Plunien}},
  \bibinfo{author}{\bibfnamefont{V.~M.} \bibnamefont{Shabaev}},
  \bibnamefont{and} \bibinfo{author}{\bibfnamefont{G.}~\bibnamefont{Soff}},
  \bibinfo{journal}{Phys. Rev. A} \textbf{\bibinfo{volume}{60}},
  \bibinfo{pages}{3522} (\bibinfo{year}{1999}).

\bibitem[{\citenamefont{Labzowsky et~al.}(1999)\citenamefont{Labzowsky,
  Goidenko, Tokman, and Pyykk{\"o}}}]{labzowsky:1999:2707}
\bibinfo{author}{\bibfnamefont{L.}~\bibnamefont{Labzowsky}},
  \bibinfo{author}{\bibfnamefont{I.}~\bibnamefont{Goidenko}},
  \bibinfo{author}{\bibfnamefont{M.}~\bibnamefont{Tokman}}, \bibnamefont{and}
  \bibinfo{author}{\bibfnamefont{P.}~\bibnamefont{Pyykk{\"o}}},
  \bibinfo{journal}{Phys. Rev. A} \textbf{\bibinfo{volume}{59}},
  \bibinfo{pages}{2707} (\bibinfo{year}{1999}).

\bibitem[{\citenamefont{Yerokhin et~al.}(2003)\citenamefont{Yerokhin,
  Indelicato, and Shabaev}}]{yerokhin:2003:073001}
\bibinfo{author}{\bibfnamefont{V.~A.} \bibnamefont{Yerokhin}},
  \bibinfo{author}{\bibfnamefont{P.}~\bibnamefont{Indelicato}},
  \bibnamefont{and} \bibinfo{author}{\bibfnamefont{V.~M.}
  \bibnamefont{Shabaev}}, \bibinfo{journal}{Phys. Rev. Lett.}
  \textbf{\bibinfo{volume}{91}}, \bibinfo{pages}{073001}
  (\bibinfo{year}{2003}).

\bibitem[{\citenamefont{Artemyev et~al.}(2005)\citenamefont{Artemyev, Shabaev,
  Yerokhin, Plunien, and Soff}}]{artemyev:2005:062104}
\bibinfo{author}{\bibfnamefont{A.~N.} \bibnamefont{Artemyev}},
  \bibinfo{author}{\bibfnamefont{V.~M.} \bibnamefont{Shabaev}},
  \bibinfo{author}{\bibfnamefont{V.~A.} \bibnamefont{Yerokhin}},
  \bibinfo{author}{\bibfnamefont{G.}~\bibnamefont{Plunien}}, \bibnamefont{and}
  \bibinfo{author}{\bibfnamefont{G.}~\bibnamefont{Soff}},
  \bibinfo{journal}{Phys. Rev. A} \textbf{\bibinfo{volume}{71}},
  \bibinfo{pages}{062104} (\bibinfo{year}{2005}).

\bibitem[{\citenamefont{Artemyev et~al.}(2007)\citenamefont{Artemyev, Shabaev,
  Tupitsyn, Plunien, and Yerokhin}}]{artemyev:2007:173004}
\bibinfo{author}{\bibfnamefont{A.~N.} \bibnamefont{Artemyev}},
  \bibinfo{author}{\bibfnamefont{V.~M.} \bibnamefont{Shabaev}},
  \bibinfo{author}{\bibfnamefont{I.~I.} \bibnamefont{Tupitsyn}},
  \bibinfo{author}{\bibfnamefont{G.}~\bibnamefont{Plunien}}, \bibnamefont{and}
  \bibinfo{author}{\bibfnamefont{V.~A.} \bibnamefont{Yerokhin}},
  \bibinfo{journal}{Phys. Rev. Lett.} \textbf{\bibinfo{volume}{98}},
  \bibinfo{pages}{173004} (\bibinfo{year}{2007}).

\bibitem[{\citenamefont{Kozhedub et~al.}(2010)\citenamefont{Kozhedub, Volotka,
  Artemyev, Glazov, Plunien, Shabaev, Tupitsyn, and
  St\"ohlker}}]{kozhedub:2010:042513}
\bibinfo{author}{\bibfnamefont{Y.~S.} \bibnamefont{Kozhedub}},
  \bibinfo{author}{\bibfnamefont{A.~V.} \bibnamefont{Volotka}},
  \bibinfo{author}{\bibfnamefont{A.~N.} \bibnamefont{Artemyev}},
  \bibinfo{author}{\bibfnamefont{D.~A.} \bibnamefont{Glazov}},
  \bibinfo{author}{\bibfnamefont{G.}~\bibnamefont{Plunien}},
  \bibinfo{author}{\bibfnamefont{V.~M.} \bibnamefont{Shabaev}},
  \bibinfo{author}{\bibfnamefont{I.~I.} \bibnamefont{Tupitsyn}},
  \bibnamefont{and}
  \bibinfo{author}{\bibfnamefont{T.}~\bibnamefont{St\"ohlker}},
  \bibinfo{journal}{Phys. Rev. A} \textbf{\bibinfo{volume}{81}},
  \bibinfo{pages}{042513} (\bibinfo{year}{2010}).

\bibitem[{\citenamefont{Sapirstein and Cheng}(2011)}]{sapirstein:2011:012504}
\bibinfo{author}{\bibfnamefont{J.}~\bibnamefont{Sapirstein}} \bibnamefont{and}
  \bibinfo{author}{\bibfnamefont{K.~T.} \bibnamefont{Cheng}},
  \bibinfo{journal}{Phys. Rev. A} \textbf{\bibinfo{volume}{83}},
  \bibinfo{pages}{012504} (\bibinfo{year}{2011}).

\bibitem[{\citenamefont{Artemyev et~al.}(2013)\citenamefont{Artemyev, Shabaev,
  Tupitsyn, Plunien, Surzhykov, and Fritzsche}}]{artemyev:2013:032518}
\bibinfo{author}{\bibfnamefont{A.~N.} \bibnamefont{Artemyev}},
  \bibinfo{author}{\bibfnamefont{V.~M.} \bibnamefont{Shabaev}},
  \bibinfo{author}{\bibfnamefont{I.~I.} \bibnamefont{Tupitsyn}},
  \bibinfo{author}{\bibfnamefont{G.}~\bibnamefont{Plunien}},
  \bibinfo{author}{\bibfnamefont{A.}~\bibnamefont{Surzhykov}},
  \bibnamefont{and}
  \bibinfo{author}{\bibfnamefont{S.}~\bibnamefont{Fritzsche}},
  \bibinfo{journal}{Phys. Rev. A} \textbf{\bibinfo{volume}{88}},
  \bibinfo{pages}{032518} (\bibinfo{year}{2013}).

\bibitem[{\citenamefont{Malyshev et~al.}(2014)\citenamefont{Malyshev, Volotka,
  Glazov, Tupitsyn, Shabaev, and Plunien}}]{malyshev:2014:062517}
\bibinfo{author}{\bibfnamefont{A.~V.} \bibnamefont{Malyshev}},
  \bibinfo{author}{\bibfnamefont{A.~V.} \bibnamefont{Volotka}},
  \bibinfo{author}{\bibfnamefont{D.~A.} \bibnamefont{Glazov}},
  \bibinfo{author}{\bibfnamefont{I.~I.} \bibnamefont{Tupitsyn}},
  \bibinfo{author}{\bibfnamefont{V.~M.} \bibnamefont{Shabaev}},
  \bibnamefont{and} \bibinfo{author}{\bibfnamefont{G.}~\bibnamefont{Plunien}},
  \bibinfo{journal}{Phys. Rev. A} \textbf{\bibinfo{volume}{90}},
  \bibinfo{pages}{062517} (\bibinfo{year}{2014}).

\bibitem[{\citenamefont{Malyshev et~al.}(2015)\citenamefont{Malyshev, Volotka,
  Glazov, Tupitsyn, Shabaev, and Plunien}}]{malyshev:2015:012514}
\bibinfo{author}{\bibfnamefont{A.~V.} \bibnamefont{Malyshev}},
  \bibinfo{author}{\bibfnamefont{A.~V.} \bibnamefont{Volotka}},
  \bibinfo{author}{\bibfnamefont{D.~A.} \bibnamefont{Glazov}},
  \bibinfo{author}{\bibfnamefont{I.~I.} \bibnamefont{Tupitsyn}},
  \bibinfo{author}{\bibfnamefont{V.~M.} \bibnamefont{Shabaev}},
  \bibnamefont{and} \bibinfo{author}{\bibfnamefont{G.}~\bibnamefont{Plunien}},
  \bibinfo{journal}{Phys. Rev. A} \textbf{\bibinfo{volume}{92}},
  \bibinfo{pages}{012514} (\bibinfo{year}{2015}).

\bibitem[{\citenamefont{Yerokhin and Shabaev}(2015)}]{yerokhin:2015:033103}
\bibinfo{author}{\bibfnamefont{V.~A.} \bibnamefont{Yerokhin}} \bibnamefont{and}
  \bibinfo{author}{\bibfnamefont{V.~M.} \bibnamefont{Shabaev}},
  \bibinfo{journal}{J. Phys. Chem. Ref. Data} \textbf{\bibinfo{volume}{44}},
  \bibinfo{pages}{033103} (\bibinfo{year}{2015}).

\bibitem[{\citenamefont{Sapirstein and Cheng}(2015)}]{sapirstein:2015:062508}
\bibinfo{author}{\bibfnamefont{J.}~\bibnamefont{Sapirstein}} \bibnamefont{and}
  \bibinfo{author}{\bibfnamefont{K.~T.} \bibnamefont{Cheng}},
  \bibinfo{journal}{Phys. Rev. A} \textbf{\bibinfo{volume}{91}},
  \bibinfo{pages}{062508} (\bibinfo{year}{2015}).

\bibitem[{\citenamefont{Malyshev et~al.}(2017)\citenamefont{Malyshev, Glazov,
  Volotka, Tupitsyn, Shabaev, Plunien, and St\"ohlker}}]{malyshev:2017:022512}
\bibinfo{author}{\bibfnamefont{A.~V.} \bibnamefont{Malyshev}},
  \bibinfo{author}{\bibfnamefont{D.~A.} \bibnamefont{Glazov}},
  \bibinfo{author}{\bibfnamefont{A.~V.} \bibnamefont{Volotka}},
  \bibinfo{author}{\bibfnamefont{I.~I.} \bibnamefont{Tupitsyn}},
  \bibinfo{author}{\bibfnamefont{V.~M.} \bibnamefont{Shabaev}},
  \bibinfo{author}{\bibfnamefont{G.}~\bibnamefont{Plunien}}, \bibnamefont{and}
  \bibinfo{author}{\bibfnamefont{T.}~\bibnamefont{St\"ohlker}},
  \bibinfo{journal}{Phys. Rev. A} \textbf{\bibinfo{volume}{96}},
  \bibinfo{pages}{022512} (\bibinfo{year}{2017}).

\bibitem[{\citenamefont{Blundell}(1993)}]{blundell:1993:1790}
\bibinfo{author}{\bibfnamefont{S.~A.} \bibnamefont{Blundell}},
  \bibinfo{journal}{Phys. Rev. A} \textbf{\bibinfo{volume}{47}},
  \bibinfo{pages}{1790} (\bibinfo{year}{1993}).

\bibitem[{\citenamefont{Sapirstein and Cheng}(2002)}]{sapirstein:2002:042501}
\bibinfo{author}{\bibfnamefont{J.}~\bibnamefont{Sapirstein}} \bibnamefont{and}
  \bibinfo{author}{\bibfnamefont{K.~T.} \bibnamefont{Cheng}},
  \bibinfo{journal}{Phys. Rev. A} \textbf{\bibinfo{volume}{66}},
  \bibinfo{pages}{042501} (\bibinfo{year}{2002}).

\bibitem[{\citenamefont{Sapirstein and Cheng}(2003)}]{sapirstein:2003:042111}
\bibinfo{author}{\bibfnamefont{J.}~\bibnamefont{Sapirstein}} \bibnamefont{and}
  \bibinfo{author}{\bibfnamefont{K.~T.} \bibnamefont{Cheng}},
  \bibinfo{journal}{Phys. Rev. A} \textbf{\bibinfo{volume}{68}},
  \bibinfo{pages}{042111} (\bibinfo{year}{2003}).

\bibitem[{\citenamefont{Chen et~al.}(2006)\citenamefont{Chen, Cheng, Johnson,
  and Sapirstein}}]{chen:2006:042510}
\bibinfo{author}{\bibfnamefont{M.~H.} \bibnamefont{Chen}},
  \bibinfo{author}{\bibfnamefont{K.~T.} \bibnamefont{Cheng}},
  \bibinfo{author}{\bibfnamefont{W.~R.} \bibnamefont{Johnson}},
  \bibnamefont{and}
  \bibinfo{author}{\bibfnamefont{J.}~\bibnamefont{Sapirstein}},
  \bibinfo{journal}{Phys. Rev. A} \textbf{\bibinfo{volume}{74}},
  \bibinfo{pages}{042510} (\bibinfo{year}{2006}).

\bibitem[{\citenamefont{Welton}(1948)}]{welton:1948:1157}
\bibinfo{author}{\bibfnamefont{T.~A.} \bibnamefont{Welton}},
  \bibinfo{journal}{Phys. Rev.} \textbf{\bibinfo{volume}{74}},
  \bibinfo{pages}{1157} (\bibinfo{year}{1948}).

\bibitem[{\citenamefont{Indelicato and Desclaux}(1990)}]{indelicato:1990:5139}
\bibinfo{author}{\bibfnamefont{P.}~\bibnamefont{Indelicato}} \bibnamefont{and}
  \bibinfo{author}{\bibfnamefont{J.~P.} \bibnamefont{Desclaux}},
  \bibinfo{journal}{Phys. Rev. A} \textbf{\bibinfo{volume}{42}},
  \bibinfo{pages}{5139} (\bibinfo{year}{1990}).

\bibitem[{\citenamefont{Pyykk{\"o} and Zhao}(2003)}]{pyykko:2003:1469}
\bibinfo{author}{\bibfnamefont{P.}~\bibnamefont{Pyykk{\"o}}} \bibnamefont{and}
  \bibinfo{author}{\bibfnamefont{L.-B.} \bibnamefont{Zhao}},
  \bibinfo{journal}{J. Phys. B} \textbf{\bibinfo{volume}{36}},
  \bibinfo{pages}{1469} (\bibinfo{year}{2003}).

\bibitem[{\citenamefont{Flambaum and Ginges}(2005)}]{flambaum:2005:052115}
\bibinfo{author}{\bibfnamefont{V.~V.} \bibnamefont{Flambaum}} \bibnamefont{and}
  \bibinfo{author}{\bibfnamefont{J.~S.~M.} \bibnamefont{Ginges}},
  \bibinfo{journal}{Phys. Rev. A} \textbf{\bibinfo{volume}{72}},
  \bibinfo{pages}{052115} (\bibinfo{year}{2005}).

\bibitem[{\citenamefont{Lowe et~al.}(2013)\citenamefont{Lowe, Chantler, and
  Grant}}]{lowe:2013:118}
\bibinfo{author}{\bibfnamefont{J.~A.} \bibnamefont{Lowe}},
  \bibinfo{author}{\bibfnamefont{C.~T.} \bibnamefont{Chantler}},
  \bibnamefont{and} \bibinfo{author}{\bibfnamefont{I.~P.} \bibnamefont{Grant}},
  \bibinfo{journal}{Radiat. Phys. Chem.} \textbf{\bibinfo{volume}{85}},
  \bibinfo{pages}{118} (\bibinfo{year}{2013}).

\bibitem[{\citenamefont{Shabaev et~al.}(2013)\citenamefont{Shabaev, Tupitsyn,
  and Yerokhin}}]{shabaev:2013:012513}
\bibinfo{author}{\bibfnamefont{V.~M.} \bibnamefont{Shabaev}},
  \bibinfo{author}{\bibfnamefont{I.~I.} \bibnamefont{Tupitsyn}},
  \bibnamefont{and} \bibinfo{author}{\bibfnamefont{V.~A.}
  \bibnamefont{Yerokhin}}, \bibinfo{journal}{Phys. Rev. A}
  \textbf{\bibinfo{volume}{88}}, \bibinfo{pages}{012513}
  (\bibinfo{year}{2013}).

\bibitem[{\citenamefont{Ginges and Berengut}(2016)}]{ginges:2016:052509}
\bibinfo{author}{\bibfnamefont{J.~S.~M.} \bibnamefont{Ginges}}
  \bibnamefont{and} \bibinfo{author}{\bibfnamefont{J.~C.}
  \bibnamefont{Berengut}}, \bibinfo{journal}{Phys. Rev. A}
  \textbf{\bibinfo{volume}{93}}, \bibinfo{pages}{052509}
  (\bibinfo{year}{2016}).

\bibitem[{\citenamefont{Tupitsyn et~al.}(2016)\citenamefont{Tupitsyn, Kozlov,
  Safronova, Shabaev, and Dzuba}}]{tupitsyn:2016:253001}
\bibinfo{author}{\bibfnamefont{I.~I.} \bibnamefont{Tupitsyn}},
  \bibinfo{author}{\bibfnamefont{M.~G.} \bibnamefont{Kozlov}},
  \bibinfo{author}{\bibfnamefont{M.~S.} \bibnamefont{Safronova}},
  \bibinfo{author}{\bibfnamefont{V.~M.} \bibnamefont{Shabaev}},
  \bibnamefont{and} \bibinfo{author}{\bibfnamefont{V.~A.} \bibnamefont{Dzuba}},
  \bibinfo{journal}{Phys. Rev. Lett.} \textbf{\bibinfo{volume}{117}},
  \bibinfo{pages}{253001} (\bibinfo{year}{2016}).

\bibitem[{\citenamefont{Dragani{\'c} et~al.}(2003)\citenamefont{Dragani{\'c},
  {Crespo~L\'opez-Urrutia}, DuBois, Fritzsche, Shabaev, {Soria~Orts}, Tupitsyn,
  Zou, and Ullrich}}]{draganic:2003:183001}
\bibinfo{author}{\bibfnamefont{I.}~\bibnamefont{Dragani{\'c}}},
  \bibinfo{author}{\bibfnamefont{J.~R.}
  \bibnamefont{{Crespo~L\'opez-Urrutia}}},
  \bibinfo{author}{\bibfnamefont{R.}~\bibnamefont{DuBois}},
  \bibinfo{author}{\bibfnamefont{S.}~\bibnamefont{Fritzsche}},
  \bibinfo{author}{\bibfnamefont{V.~M.} \bibnamefont{Shabaev}},
  \bibinfo{author}{\bibfnamefont{R.}~\bibnamefont{{Soria~Orts}}},
  \bibinfo{author}{\bibfnamefont{I.~I.} \bibnamefont{Tupitsyn}},
  \bibinfo{author}{\bibfnamefont{Y.}~\bibnamefont{Zou}}, \bibnamefont{and}
  \bibinfo{author}{\bibfnamefont{J.}~\bibnamefont{Ullrich}},
  \bibinfo{journal}{Phys. Rev. Lett.} \textbf{\bibinfo{volume}{91}},
  \bibinfo{pages}{183001} (\bibinfo{year}{2003}).

\bibitem[{\citenamefont{Pa\v{s}teka et~al.}(2017)\citenamefont{Pa\v{s}teka,
  Eliav, Borschevsky, Kaldor, and Schwerdtfeger}}]{pasteka:2017:023002}
\bibinfo{author}{\bibfnamefont{L.~F.} \bibnamefont{Pa\v{s}teka}},
  \bibinfo{author}{\bibfnamefont{E.}~\bibnamefont{Eliav}},
  \bibinfo{author}{\bibfnamefont{A.}~\bibnamefont{Borschevsky}},
  \bibinfo{author}{\bibfnamefont{U.}~\bibnamefont{Kaldor}}, \bibnamefont{and}
  \bibinfo{author}{\bibfnamefont{P.}~\bibnamefont{Schwerdtfeger}},
  \bibinfo{journal}{Phys. Rev. Lett.} \textbf{\bibinfo{volume}{118}},
  \bibinfo{pages}{023002} (\bibinfo{year}{2017}).

\bibitem[{\citenamefont{Li et~al.}(2018)\citenamefont{Li, Si, Brage, Hutton,
  and Zou}}]{li:2018:R020502}
\bibinfo{author}{\bibfnamefont{M.~C.} \bibnamefont{Li}},
  \bibinfo{author}{\bibfnamefont{R.}~\bibnamefont{Si}},
  \bibinfo{author}{\bibfnamefont{T.}~\bibnamefont{Brage}},
  \bibinfo{author}{\bibfnamefont{R.}~\bibnamefont{Hutton}}, \bibnamefont{and}
  \bibinfo{author}{\bibfnamefont{Y.~M.} \bibnamefont{Zou}},
  \bibinfo{journal}{Phys. Rev. A} \textbf{\bibinfo{volume}{98}},
  \bibinfo{pages}{020502(R)} (\bibinfo{year}{2018}).

\bibitem[{\citenamefont{Si et~al.}(2018)\citenamefont{Si, Guo, Brage, Chen,
  Hutton, and {Froese Fischer}}}]{si:2018:012504}
\bibinfo{author}{\bibfnamefont{R.}~\bibnamefont{Si}},
  \bibinfo{author}{\bibfnamefont{X.~L.} \bibnamefont{Guo}},
  \bibinfo{author}{\bibfnamefont{T.}~\bibnamefont{Brage}},
  \bibinfo{author}{\bibfnamefont{C.~Y.} \bibnamefont{Chen}},
  \bibinfo{author}{\bibfnamefont{R.}~\bibnamefont{Hutton}}, \bibnamefont{and}
  \bibinfo{author}{\bibfnamefont{C.}~\bibnamefont{{Froese Fischer}}},
  \bibinfo{journal}{Phys. Rev. A} \textbf{\bibinfo{volume}{98}},
  \bibinfo{pages}{012504} (\bibinfo{year}{2018}).

\bibitem[{\citenamefont{Layzer}(1959)}]{layzer:1959:271}
\bibinfo{author}{\bibfnamefont{D.}~\bibnamefont{Layzer}},
  \bibinfo{journal}{Ann. Phys. (NY)} \textbf{\bibinfo{volume}{8}},
  \bibinfo{pages}{271} (\bibinfo{year}{1959}).

\bibitem[{\citenamefont{Layzer and Bahcall}(1962)}]{layzer:1962:177}
\bibinfo{author}{\bibfnamefont{D.}~\bibnamefont{Layzer}} \bibnamefont{and}
  \bibinfo{author}{\bibfnamefont{J.}~\bibnamefont{Bahcall}},
  \bibinfo{journal}{Ann. Phys. (NY)} \textbf{\bibinfo{volume}{17}},
  \bibinfo{pages}{177} (\bibinfo{year}{1962}).

\bibitem[{\citenamefont{Beiersdorfer et~al.}(1998)\citenamefont{Beiersdorfer,
  Osterheld, and Elliott}}]{beiersdorfer:1998:1944}
\bibinfo{author}{\bibfnamefont{P.}~\bibnamefont{Beiersdorfer}},
  \bibinfo{author}{\bibfnamefont{A.~L.} \bibnamefont{Osterheld}},
  \bibnamefont{and} \bibinfo{author}{\bibfnamefont{S.~R.}
  \bibnamefont{Elliott}}, \bibinfo{journal}{Phys. Rev. A}
  \textbf{\bibinfo{volume}{58}}, \bibinfo{pages}{1944} (\bibinfo{year}{1998}).

\bibitem[{\citenamefont{Shabaev et~al.}(2004)\citenamefont{Shabaev, Tupitsyn,
  Yerokhin, Plunien, and Soff}}]{shabaev:2004:130405}
\bibinfo{author}{\bibfnamefont{V.~M.} \bibnamefont{Shabaev}},
  \bibinfo{author}{\bibfnamefont{I.~I.} \bibnamefont{Tupitsyn}},
  \bibinfo{author}{\bibfnamefont{V.~A.} \bibnamefont{Yerokhin}},
  \bibinfo{author}{\bibfnamefont{G.}~\bibnamefont{Plunien}}, \bibnamefont{and}
  \bibinfo{author}{\bibfnamefont{G.}~\bibnamefont{Soff}},
  \bibinfo{journal}{Phys. Rev. Lett.} \textbf{\bibinfo{volume}{93}},
  \bibinfo{pages}{130405} (\bibinfo{year}{2004}).

\bibitem[{\citenamefont{Sapirstein and Johnson}(1996)}]{sapirstein:1996:5213}
\bibinfo{author}{\bibfnamefont{J.}~\bibnamefont{Sapirstein}} \bibnamefont{and}
  \bibinfo{author}{\bibfnamefont{W.~R.} \bibnamefont{Johnson}},
  \bibinfo{journal}{J. Phys. B} \textbf{\bibinfo{volume}{29}},
  \bibinfo{pages}{5213} (\bibinfo{year}{1996}).

\bibitem[{\citenamefont{J{\"o}nsson et~al.}(2013)\citenamefont{J{\"o}nsson,
  Gaigalas, Biero\'n, {Froese Fischer}, and Grant}}]{joensson:2013:2197}
\bibinfo{author}{\bibfnamefont{P.}~\bibnamefont{J{\"o}nsson}},
  \bibinfo{author}{\bibfnamefont{G.}~\bibnamefont{Gaigalas}},
  \bibinfo{author}{\bibfnamefont{J.}~\bibnamefont{Biero\'n}},
  \bibinfo{author}{\bibfnamefont{C.}~\bibnamefont{{Froese Fischer}}},
  \bibnamefont{and} \bibinfo{author}{\bibfnamefont{I.~P.} \bibnamefont{Grant}},
  \bibinfo{journal}{Comput. Phys. Commun.} \textbf{\bibinfo{volume}{184}},
  \bibinfo{pages}{2197} (\bibinfo{year}{2013}).

\bibitem[{\citenamefont{Kohn and Sham}(1965)}]{kohn:1965:A1133}
\bibinfo{author}{\bibfnamefont{W.}~\bibnamefont{Kohn}} \bibnamefont{and}
  \bibinfo{author}{\bibfnamefont{L.~J.} \bibnamefont{Sham}},
  \bibinfo{journal}{Phys. Rev.} \textbf{\bibinfo{volume}{140}},
  \bibinfo{pages}{A1133} (\bibinfo{year}{1965}).

\bibitem[{\citenamefont{Peacock et~al.}(1984)\citenamefont{Peacock, Stamp, and
  Silver}}]{peacock:1984:1984}
\bibinfo{author}{\bibfnamefont{N.~J.} \bibnamefont{Peacock}},
  \bibinfo{author}{\bibfnamefont{M.~F.} \bibnamefont{Stamp}}, \bibnamefont{and}
  \bibinfo{author}{\bibfnamefont{J.~D.} \bibnamefont{Silver}},
  \bibinfo{journal}{Phys. Scr.} \textbf{\bibinfo{volume}{1984}},
  \bibinfo{pages}{10} (\bibinfo{year}{1984}).

\bibitem[{\citenamefont{Denne et~al.}(1989)\citenamefont{Denne, Hinnov,
  Ramette, and Saoutic}}]{denne:1989:1488}
\bibinfo{author}{\bibfnamefont{B.}~\bibnamefont{Denne}},
  \bibinfo{author}{\bibfnamefont{E.}~\bibnamefont{Hinnov}},
  \bibinfo{author}{\bibfnamefont{J.}~\bibnamefont{Ramette}}, \bibnamefont{and}
  \bibinfo{author}{\bibfnamefont{B.}~\bibnamefont{Saoutic}},
  \bibinfo{journal}{Phys. Rev. A} \textbf{\bibinfo{volume}{40}},
  \bibinfo{pages}{1488} (\bibinfo{year}{1989}).

\bibitem[{\citenamefont{Silwal et~al.}(2017)\citenamefont{Silwal, Takacs,
  Dreiling, Gillaspy, and Ralchenko}}]{silwal:2017:30}
\bibinfo{author}{\bibfnamefont{R.}~\bibnamefont{Silwal}},
  \bibinfo{author}{\bibfnamefont{E.}~\bibnamefont{Takacs}},
  \bibinfo{author}{\bibfnamefont{J.~M.} \bibnamefont{Dreiling}},
  \bibinfo{author}{\bibfnamefont{J.~D.} \bibnamefont{Gillaspy}},
  \bibnamefont{and}
  \bibinfo{author}{\bibfnamefont{Y.}~\bibnamefont{Ralchenko}},
  \bibinfo{journal}{Atoms} \textbf{\bibinfo{volume}{5}}, \bibinfo{pages}{30}
  (\bibinfo{year}{2017}).

\bibitem[{\citenamefont{Myrn\"as et~al.}(1994)\citenamefont{Myrn\"as, Jup\'en,
  Miecznik, Martinson, and {Denne-Hinnov}}}]{myrnaes:1994:429}
\bibinfo{author}{\bibfnamefont{R.}~\bibnamefont{Myrn\"as}},
  \bibinfo{author}{\bibfnamefont{C.}~\bibnamefont{Jup\'en}},
  \bibinfo{author}{\bibfnamefont{G.}~\bibnamefont{Miecznik}},
  \bibinfo{author}{\bibfnamefont{I.}~\bibnamefont{Martinson}},
  \bibnamefont{and}
  \bibinfo{author}{\bibfnamefont{B.}~\bibnamefont{{Denne-Hinnov}}},
  \bibinfo{journal}{Phys. Scr.} \textbf{\bibinfo{volume}{49}},
  \bibinfo{pages}{429} (\bibinfo{year}{1994}).

\bibitem[{\citenamefont{Clementson et~al.}(2011)\citenamefont{Clementson,
  Beiersdorfer, Brown, Gu, Lundberg, Podpaly, and
  Tr\"abert}}]{clementson:2011:571}
\bibinfo{author}{\bibfnamefont{J.}~\bibnamefont{Clementson}},
  \bibinfo{author}{\bibfnamefont{P.}~\bibnamefont{Beiersdorfer}},
  \bibinfo{author}{\bibfnamefont{G.~V.} \bibnamefont{Brown}},
  \bibinfo{author}{\bibfnamefont{M.~F.} \bibnamefont{Gu}},
  \bibinfo{author}{\bibfnamefont{H.}~\bibnamefont{Lundberg}},
  \bibinfo{author}{\bibfnamefont{Y.}~\bibnamefont{Podpaly}}, \bibnamefont{and}
  \bibinfo{author}{\bibfnamefont{E.}~\bibnamefont{Tr\"abert}},
  \bibinfo{journal}{Can. J. Phys.} \textbf{\bibinfo{volume}{89}},
  \bibinfo{pages}{571} (\bibinfo{year}{2011}).

\bibitem[{\citenamefont{Mohr and Sapirstein}(2000)}]{mohr:2000:052501}
\bibinfo{author}{\bibfnamefont{P.~J.} \bibnamefont{Mohr}} \bibnamefont{and}
  \bibinfo{author}{\bibfnamefont{J.}~\bibnamefont{Sapirstein}},
  \bibinfo{journal}{Phys. Rev. A} \textbf{\bibinfo{volume}{62}},
  \bibinfo{pages}{052501} (\bibinfo{year}{2000}).

\bibitem[{\citenamefont{Yerokhin et~al.}(2001)\citenamefont{Yerokhin, Artemyev,
  Shabaev, Sysak, Zherebtsov, and Soff}}]{yerokhin:2001:032109}
\bibinfo{author}{\bibfnamefont{V.~A.} \bibnamefont{Yerokhin}},
  \bibinfo{author}{\bibfnamefont{A.~N.} \bibnamefont{Artemyev}},
  \bibinfo{author}{\bibfnamefont{V.~M.} \bibnamefont{Shabaev}},
  \bibinfo{author}{\bibfnamefont{M.~M.} \bibnamefont{Sysak}},
  \bibinfo{author}{\bibfnamefont{O.~M.} \bibnamefont{Zherebtsov}},
  \bibnamefont{and} \bibinfo{author}{\bibfnamefont{G.}~\bibnamefont{Soff}},
  \bibinfo{journal}{Phys. Rev. A} \textbf{\bibinfo{volume}{64}},
  \bibinfo{pages}{032109} (\bibinfo{year}{2001}).

\end{thebibliography}
\end{document}